# Vibration spectra of benzene-like models with Hooke's law interactions

M. M. Bogdan[1], O. V. Charkina[1,2], and A. Y. Holovashchenko[3]

[1]*B. Verkin Institute for Low Temperature Physics and Engineering of the National Academy of Sciences of Ukraine, 47 Nauky Ave., Kharkiv, 61103, Ukraine*

[2] *University of Luxembourg, L-1511 Luxembourg City, Luxembourg*

[3]*V.N. Karazin Kharkiv National University, 4 Svobody Sq., Kharkiv, 610022, Ukraine*

E-mail: charkina@ilt.kharkov.ua
m_m_bogdan@ukr.net

## Abstract

The harmonic oscillations of a spring-ball model of benzene-like nanosystems with Hooke's law interactions between nearest, second, and third neighbors are explored. We show that in the cylindrical coordinates the dynamics of this cyclic hexagonal system is described by the Lagrange equations similar to those of the one-dimensional two-component crystal model. We expose that the vibration frequencies of the hexagonal model lie on the branches of the dispersion law of the associated lattice model, and their positions are determined by the cyclic Born-Von Karman condition. The hexagonal model is generalized to one describing the benzene molecule and the fully deuterated and halogenated benzenes. The effect of hybridization of vibration modes and pushing apart of spectral branches in the crossover situation is revealed. All the discrete frequency spectrum and normal modes of oscillations and their explicit dependencies on all the constants of elastic interactions are exactly found.

## Introduction

Dynamical and thermodynamical properties of carbon-based compounds are subjects of frontline research in low-temperature physics. Academician V.G. Manzhelii was a creator of the scientific school of investigations of thermal properties of cryogenic crystals. At present, following novel trends, his scholars study thermoconductive and spectroscopic properties of complex molecular compounds, which include now modern nanostructures such as carbon nanotubes, one-dimensional atomic chains in the grooves of carbon nanotube bundles, fullerites and fullerenes, and other carbon-based molecules as impurities in polyatomic lattices. Remembering with deep respect V.G. Manzhelii as a prominent scientist and a great teacher in life, we dedicate this paper to his memory on the 90th anniversary of his birth.

The investigation of the structure and spectral properties of polyatomic molecules is one of the central tasks of molecular physics [1]. The vibrational spectra of such molecules contain significant information about their structural and dynamic properties. Since the discovery and subsequent development of the methods of infrared and Raman spectroscopy [2], the analysis of the vibration spectra of polyatomic molecules has been one of the main methods of determining the physical and chemical properties of matter. The study of the features of the spectrum of vibration frequencies of a single simple molecule or a group of atoms in a complex molecule makes it possible to identify trends in the chemical structure of compounds and to serve in the future for the development of technologies for the synthesis of new materials with predetermined properties.

The basis for understanding the dynamic behavior of molecules is classical and quantum mechanics. In the approximation of harmonic oscillations, the dynamics of a finite system of interacting atoms is reduced in the basis of normal modes to the dynamics of an ensemble of independent oscillators [3]. As a result of solving the classical equations, the determined discrete spectrum of frequencies $\omega_i$ is the main dynamic characteristic of such a set of oscillators. Quantum theory gives for each $i$-th oscillator an equidistant spectrum of energy levels with the energy quantum $\hbar\omega_i$. This makes it possible to build a theory of the molecule interaction with electromagnetic waves as the basis of spectroscopy methods, as well as to calculate the contribution of internal vibrations of molecules to their thermodynamics.

As known [3], the foundations of the classical theory of molecular oscillations were established in the 1930s, in particular in the works of E. Bright Wilson [4,5]. Although the problem of the dynamics of a system of interacting oscillators is reduced, from the mathematical point of view, to a spectral problem of finding eigenvectors and eigenvalues, the characteristic equations for the oscillation frequencies, except for a few cases of the simplest molecules, cannot be solved explicitly. Because of this, Wilson's approach was based on taking into account the possible elements of symmetry of molecules and the group theory, which would allow to identify the main types of normal modes of oscillations and to obtain for them simplified expressions of the characteristic equation for the simplest models of interatomic interaction.

Due to the lack of explicit dependences of vibration frequencies on the parameters of molecular models, the results of dynamic experiments with mechanical models, which are built from metallic balls connected by springs, were used for the analysis of infrared and Raman spectroscopy data [6,7]. It has been interesting that such mechanical models turned out to be effective not only for the qualitative but also for the quantitative description of the spectra of molecules.

Further progress in the classical description of molecular dynamics consisted in improving the choice of variables as so-called internal collective coordinates of atoms in molecules, taking into account their symmetry [8]. But with the development of computing capabilities, from simple computers to supercomputers, spectral problems of mechanical vibrations began to be solved numerically for each individual molecule with a given set of its parameters. Finally, quantum mechanical methods were invented and developed, in particular, the method of density functional theory (DFT) [9], within which the electron density distribution, the interaction of atoms, and, ultimately, the electronic and vibrational spectra of molecules are calculated from first principles.

But the general classification of the types of molecular vibrations is based rather on the determination of their normal modes. Among them, one distinguishes the valence modes caused by a change in bond lengths, and the deformation modes of orientational vibrations, and active and inactive modes with respect to the change in the dipole moment. The frequencies are divided into two groups of the infrared and Raman bands. In addition, the vibration frequencies of individual atoms or groups of atoms are also distinguished. In any case, finally, the classification of frequencies is characterized by their quantitative values.

The main difficulty of the theory of vibrations of most molecules is related to the fact that the spectral problem for them has free boundary conditions due to the free ends of molecules, which makes it impossible to find an exact general analytical form of frequencies and normal modes of vibrations.

In the present work, we show that the spectral problem on normal vibration modes and eigenfrequencies is solved exactly for cyclic molecular structures, such as benzene-like molecules. Due to the structure periodicity, the theory of the crystal lattice can be applied to such systems [10]. The main theoretical calculation idea of the paper is the use of cylindrical (polar) coordinates as a new approach to the analytical description of the dynamics of cyclic molecules. The proposed approach is demonstrated on a mechanical model of six balls, all of which are connected to each other by springs with different stiffness coefficients depending on the particle spacing. We found the explicit form of normal modes and all vibration frequencies for such a hexagonal elastic system. We showed that they belong to the two branches of the dispersion law of the associated one-dimensional two-component lattice model. Bearing in mind the closeness of the model to the benzene molecule and its related molecules, we demonstrated how the presence of hydrogen in the benzene molecule can be taken into account in the framework of the mechanical model and how the discrete frequency spectrum is related to branches of the dispersion law of the associated complex lattice. Finally, the discrete frequencies are found analytically for model parameters corresponding to the fully deuterated benzene molecule and the benzene with substitution of hydrogen by fluorine, chlorine, bromine, and iodine, respectively. At the end, it is shown how the proposed approach can be generalized to other cyclic finite particle structures of elastic origin.

**The spring-ball model of a carbon hexagon**

Current models of the benzene molecule are built on the basis of results obtained using quantum mechanical methods such as a method of DFT [9]. These studies proved that the chemical bonds and interactions between the nearest carbons in the molecule are the same throughout the ring. Thereby, they refined the previously adopted Kekule model with alternating single and double bonds [3]. Moreover, since the carbon bonds form a homogeneous ring, in addition to the interaction of the nearest neighbors, the interaction of each carbon with the second and third neighbors should be taken into account. As mentioned above, in addition to molecules originated from benzene, carbon hexagons are the main element of carbon nanotubes and two-dimensional graphene crystals. The standard set of interactions in the hexagon in the latter case consists of pair interactions between all the atoms with a strength hierarchy from the strong nearest neighbors interaction to smaller ones with the second and third neighbors. [11].

Note that these interactions have a central character, i.e., depend only on the distance between atoms [3, 10]. In the harmonic approximation, when the atoms in the hexagon oscillate with small amplitudes, forces between them arise in accordance with Hooke's law. Therefore, regardless of the origin of the central interaction, whether it is covalent bonds, or van der Waals forces, or a set of complex interactions, the potential energy will quadratically depend on the small displacements of atoms with respect to each other. That is the same as the potential energy of a spring connecting two balls in a mechanical model.

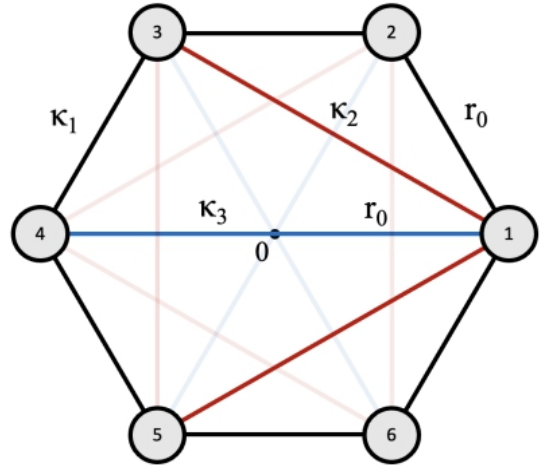

*Fig.* 1. The chart of the hexagonal mechanical model of balls connected by springs. The springs connecting the nearest neighbors are marked as the black lines, the second neighbors' springs are shown as the red lines, and the third as the blue lines. Each atom interacts with all others like it is shown for the first atom.

However, in the case of central forces, the mechanical model with the interaction of only nearest neighbors with a stiffness coefficient $\kappa_1$ is unstable since such a system of balls can be folded into a linear structure without stretching the springs with an increase in the distance between two opposite atoms and the pairwise approach of the remaining four atoms. In order to stabilize the hexagon, it is necessary to introduce for each atom an interaction with its second neighbors through additional springs with stiffness coefficient $\kappa_2$ or immediately with third ones with stiffness coefficient $\kappa_3$, as was done in the first mechanical experiments [6,7]. Therefore, for the carbon hexagon of the benzene molecule, it is natural to propose a mechanical model of balls with the same mass $m$, which are connected to the nearest, second, and third neighbors by springs with the stiffness coefficients $\kappa_1$, $\kappa_2$ and $\kappa_3$, respectively, as shown in Fig. 1. Thus in present work we consider the mechanical model with Hooke's law pair interactions of all the particles and do not include any special non-central interactions.

The benzene molecule surely contains hydrogen atoms, each of which interacts with one carbon to form a pair (see Fig. 2). The corresponding more complex elastic model, which describes the features of the dynamic behavior of benzene $C_6H_6$, fully deuterated benzene $C_6D_6$ and fully substituted benzene with fluorine, chloride, bromine, and iodine, will be formulated and considered in detail in the second part of the paper.

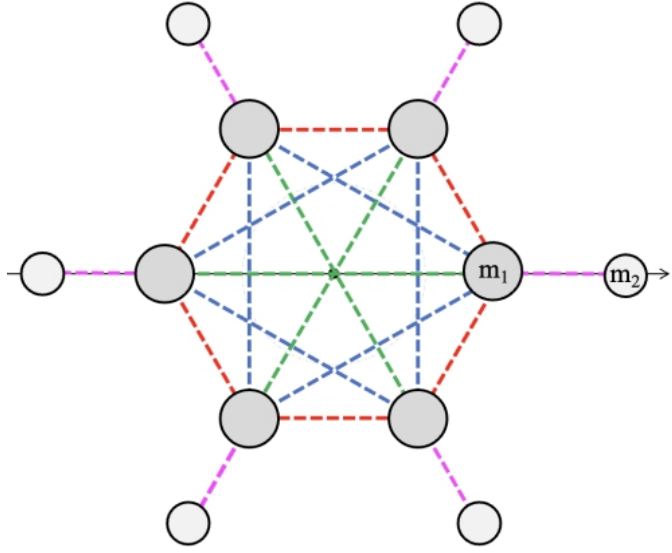

*Fig*. 2. The chart of the mechanical benzene-like model with the colored dash lines showing four types of spring bonds between particles.

In order to describe analytically the hexagonal model of particles lying in the plane (Fig. 1), we introduce cylindrical coordinates with the origin in the center of mass of the system and denote the radius vector of every particle as $\mathbf{r}_n$. Then the $n$-th particle has its cylindrical coordinates as the radius $r_n$ and the azimuth angle $\varphi_n$ in the plane and the projection $z_n$ on the axis perpendicular to the plane. In the state of rest, the particles have the following equilibrium values of the coordinates:

$$r_n^{(0)} = r_0, \qquad \varphi_n^{(0)} \equiv \frac{\pi}{3} \cdot (n-1), \qquad z_n^{(0)} = 0, \tag{1}$$

where $n = 1,..,6$ as indicated in Fig. 1. Thus, in equilibrium, each atom is at the same distance $r_0$ from each other and from the center and is deflected from each other by the angle $\pi/3$.

When particles oscillate with small amplitudes, their polar coordinates get the displacements $u_n \ll r_0$ and $\theta_n \ll 1$, as shown in Fig. 3:

$$r_n = r_0 + u_n, \qquad \varphi_n = \frac{\pi}{3}(n-1) + \theta_n \tag{2}$$

and, in general, the projection $z_n$ could be non-zero.

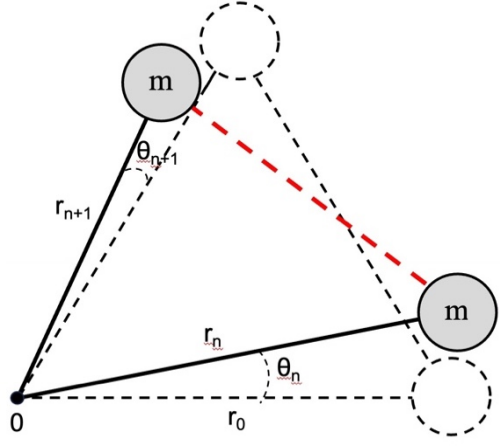

*Fig.* 3. Displacements of neighbor particles from their equilibrium positions in polar coordinates during small vibrations.

It is easy to find the Lagrangian $L = T - U$ of the system in the harmonic approximation. The kinetic energy of the system is written as follows:

$$T = \frac{m}{2} \sum_{n=1}^{6}\left[\left(\frac{du_n}{dt}\right)^2 + \left(\frac{dz_n}{dt}\right)^2\right] + \frac{I}{2}\sum_{n=1}^{6}\left(\frac{d\theta_n}{dt}\right)^2 \qquad (3)$$

where the first term is the contribution of the deviation of the length of the radius vector of the particle from $r_0$, the second is the contribution of the deviation from the plane, and the third is the contribution of the deviation of the particle azimuthal angle from the equilibrium value. The moment of inertia is equal to $I = mr_0^2$.

The potential energy $U$ is the sum of quadratic functions of *small changes in the lengths of all springs* during displacements of each oscillating atom from its equilibrium position. Since there are three types of springs in the system, the potential energy has three components:

$$U = U_1 + U_2 + U_3 = \frac{\kappa_1}{2}\sum_{n=1}^{6}\xi_n^2 + \frac{\kappa_2}{2}\sum_{n=1}^{6}\zeta_n^2 + \frac{\kappa_3}{2}\sum_{n=1}^{3}\eta_n^2 \qquad (4)$$

where $U_1$ is the potential energy of the interaction of the particles with the nearest neighbors and $\xi_n$ is a small change in the length of the first-type spring between the balls with the numbers $n$ and $n+1$, $U_2$ is the potential energy of the interaction of the balls with the second neighbors (through one) and $\zeta_n$ is the change in the length of the second-type spring between the balls with the numbers $n$ and $n+2$, and $U_3$ is the potential energy of the interaction of atoms with third neighbors (located opposite) and $\eta_n$ is the change in the length of the spring of the third type between the balls with the numbers $n$ and $n+3$. Note that there are only three springs of the third type. However, further, it is convenient to use the third sum with six terms and to divide the result of summing by two. Surely, the length changes $\eta_{n+3} \equiv \eta_n$.

The method of calculating the change in the spring length in a linear chain of atoms with central and non-central interactions in 3D space was specified in Ref. 10. This suggests a way to calculate the change in the spring length for the present hexagonal model. As seen in Fig. 3, the length of the spring, at the ends of which there are balls with the numbers $n$ and $n+1$, is equal to the vector modulus of the difference of the radius vectors of these particles. From geometric consideration, with using Pythagoras' theorem, it is easy to find that the $z$-projections of the radius vectors on the $z$-axis give a nonlinear (quadratic) contribution to the length of the spring. The matter of the fact is the zero equilibrium value of this variable in contrast to the non-zero equilibrium values of the radial and angle coordinates $r_n$ and $\varphi_n$. As a result, the contributions of displacements along the radius and deviations from the azimuthal angle to the spring length are linear. In fact, according to the law of cosines, if two sides of a triangle are equal to $r_n$ and $r_{n+1}$, and the angle between them is $\varphi_{n+1} - \varphi_n$, then its third side (the red dashed line in Fig. 3) is determined by the formula

$$l_{n,n+1} = \sqrt{r_{n+1}^2 + r_n^2 - 2r_{n+1}r_n \cos(\varphi_{n+1} - \varphi_n)} \tag{5}$$

At equilibrium values $r_n = r_{n+1} = r_0$ and $\varphi_{n+1} - \varphi_n = \pi/3$ the equilibrium length of the spring or the distance between the balls is naturally $l_{n,n+1}^{(0)} = r_0$. Now substituting the polar coordinates of the balls with displacements (2) into formula (5), we obtain

$$l_{n,n+1} = r_0 \sqrt{(1+v_{n+1})^2 + (1+v_n)^2 - 2(1+v_{n+1})(1+v_n)\cos\left(\frac{\pi}{3} + \theta_{n+1} - \theta_n\right)} \tag{6}$$

where a dimensionless displacement variable $v_n \equiv u_n/r_0 \ll 1$ is introduced. Using smallness of $v_n$ and $\theta_n$ we find approximately

$$\cos\left(\frac{\pi}{3} + \theta_{n+1} - \theta_n\right) \approx \cos\frac{\pi}{3} - \sin\frac{\pi}{3}\cdot(\theta_{n+1} - \theta_n) = \frac{1}{2}\left[1 - \sqrt{3}(\theta_{n+1} - \theta_n)\right], \tag{7}$$

and finally, we have the change of the spring length as follows

$$\xi_n = l_{n,n+1} - r_0 = \frac{r_0}{2}\left(v_{n+1} + v_n + \sqrt{3}\theta_{n+1} - \sqrt{3}\theta_n\right). \tag{8}$$

Thus, the change in the length of the spring is determined by the linear terms of the dimensionless displacements of the radial and angular coordinates of the particles at the edges of the spring.

It is easy to notice that a general formula for the change of the $j$-th spring length between $n$-th and $(n+j)$-th particles in the mechanical hexagon is as follows

$$l_{n,n+j} - l_{n,n+j}^{(0)} = \sqrt{r_{n+j}^2 + r_n^2 - 2r_{n+1}r_n \cos(\frac{j\pi}{3} + \theta_{n+j} - \theta_n)} - l_{n,n+j}^{(0)}, \quad j = 1,2,3. \tag{9}$$

Therefore the changes of the lengths of springs of the second and third types are calculated similarly to the first case. In particular, we find the following change in the second-type string length

$$\zeta_n = l_{n,n+2} - \sqrt{3}r_0 \approx \frac{r_0}{2}\left[\sqrt{3}(v_{n+2} + v_n) + \theta_{n+2} - \theta_n\right], \tag{10}$$

and the corresponding expression for the third-type string length

$$\eta_n \approx l_{n,n+3} - 2r_0 = r_0(v_{n+3} + v_n). \tag{11}$$

By substituting the Eqs. (8), (9), and (11) in the formula (4), we finally obtain the total potential energy.

We note that the dynamic problem is solved in the flat geometry because the deviation from the plane of the molecule is a nonlinear effect on the variable $z_n$, and in order to study the dynamics with taking into account the contribution of this variable to the kinetic energy, it will be necessary to go beyond the harmonic approximation.

Now we are able to write the Lagrangian for the hexagonal model of particles interacting according to Hooke's law between the first, second, and third neighbors in the harmonic approximation:

$$L = \frac{I}{2} \sum_{n=1}^{6} \left[ \left( \frac{dv_n}{dt} \right)^2 + \left( \frac{d\theta_n}{dt} \right)^2 \right] - \frac{\kappa_1 r_0^2}{8} \sum_{n=1}^{6} (v_{n+1} + v_n + \sqrt{3}\theta_{n+1} - \sqrt{3}\theta_n)^2 -$$
$$- \frac{\kappa_2 r_0^2}{24} \sum_{n=1}^{6} (3v_{n+2} + 3v_n + \sqrt{3}\theta_{n+2} - \sqrt{3}\theta_n)^2 - \frac{\kappa_3 r_0^2}{4} \sum_{n=1}^{6} (v_{n+3} + v_n)^2. \tag{12}$$

It is possible to simplify the Lagrangian form by introducing the characteristic frequency and energy, and the dimensionless time and the dimensionless stiffness parameters as follows:

$$\omega_0^2 = \frac{\kappa_1}{4m}, \quad E_0 = mr_0^2 \omega_0^2, \quad \tau = \omega_0 t, \quad \Lambda = \frac{\kappa_2}{3\kappa_1}, \quad S = \frac{2\kappa_3}{\kappa_1}. \tag{13}$$

Then the Lagrangian of the system takes the final form:

$$L = \frac{E_0}{2} \left( \sum_{n=1}^{6} \left( \frac{dv_n}{d\tau} \right)^2 + \sum_{n=1}^{6} \left( \frac{d\theta_n}{d\tau} \right)^2 - \sum_{n=1}^{6} (v_{n+1} + v_n + \sqrt{3}\theta_{n+1} - \sqrt{3}\theta_n)^2 - \right.$$
$$\left. - \Lambda \sum_{n=1}^{6} (3v_{n+2} + 3v_n + \sqrt{3}\theta_{n+2} - \sqrt{3}\theta_n)^2 - S \sum_{n=1}^{6} (v_{n+3} + v_n)^2 \right) \tag{14}$$

The Lagrangian (14) corresponds to the one-dimensional two-component model, which has six particles and evident rotational symmetry. If the radial positions of particles do not change, the system degenerates to the scalar model of the one-dimensional crystal with displacements obeying the cyclic Born-Von Karman boundary condition [10]. If azimuth angles of particles do not change, then the Lagrangian dynamics reduces to radially symmetric oscillation motion of particles of the ring system. However, as seen from Eq. (14), the interaction between these two subsystems is strong, and the total system dynamics are expected to be rather complex. Nevertheless, the dynamical problem of the hexagonal model is solved completely in the next section.

## Equations of motion and vibration spectrum of hexagonal model

The application of the methods of the crystal lattice theory [10] makes it possible to exactly solve the problem of the dynamics of the system with Lagrangian (14). The Lagrange equations for generalized coordinates $q_i$ are derived from the Lagrangian by taking the corresponding derivatives

$$\frac{d}{d\tau}\left(\frac{\partial L}{\partial \dot{q}_i}\right) - \frac{\partial L}{\partial q_i} = 0. \tag{15}$$

For the hexagonal model, the set of generalized coordinates is as follows:

$$\mathbf{q}(\tau) = (q_1, ..., q_{12}) \equiv (v_1, ... v_6, \theta_1, ... \theta_6), \tag{16}$$

where the column vector $\mathbf{q}(\tau)$, for convenience, is written as a row.

Since the kinetic energy of the system is the diagonal quadratic form of the velocities, the Lagrange equations are reduced to the following:

$$-\frac{d^2 v_n}{d\tau^2} = \frac{1}{E_0}\frac{\partial U}{\partial v_n}, \quad -\frac{d^2 \theta_n}{d\tau^2} = \frac{1}{E_0}\frac{\partial U}{\partial \theta_n}. \tag{17}$$

Taking the derivatives of the potential energy with respect to the variables $v_n$ and $\theta_n$, we find the explicit expressions of the twelve linear differential Lagrange equations. We seek for stationary solutions of these equations in the form of harmonic oscillations of the system:

$$v_n = V_n e^{-i\Omega\tau}, \quad \theta_n = \vartheta_n e^{-i\Omega\tau}, \tag{18}$$

where the parameter $\Omega$ is the dimensionless frequency. After the substitution (18) in the Lagrange equations, we reduce the dynamical problem to the spectral problem of finding the eigenvalues $\Omega^2$ of a system of homogeneous algebraic equations with constant coefficients

$$\begin{aligned}
\Omega^2 V_n &= 2V_n + V_{n+1} + V_{n-1} + \sqrt{3}(\vartheta_{n+1} - \vartheta_{n-1}) + \\
&+ 3\Lambda\big(6V_n + 3V_{n+2} + 3V_{n-2} + \sqrt{3}(\vartheta_{n+2} - \vartheta_{n-2})\big) + \\
&+ S(2V_n + V_{n+3} + V_{n-3}) \\
\Omega^2 \vartheta_n &= 6\vartheta_n - 3\vartheta_{n+1} - 3\vartheta_{n-1} + \sqrt{3}(V_{n-1} - V_{n+1}) + \\
&+ 3\Lambda\big(2\vartheta_n - \vartheta_{n+2} - \vartheta_{n-2} + \sqrt{3}(V_{n-2} - V_{n+2})\big)
\end{aligned} \tag{19}$$

The system of Eqs. (19) can be written as the eigenvalue problem of the Hermitian operator $\hat{\mathbf{A}}\mathbf{q} = \Omega^2 \mathbf{q}$ with a matrix 12x12, whose elements are coefficients of Eqs. (19). Besides, the cyclic Born-Von Karman boundary condition is imposed on the solutions of the system of Eqs. (19). As a result, the components of the constant eigenvector $\mathbf{q}$ are found in the explicit form [10]:

$$\begin{aligned}
V_n &= V_0 \exp(iKr_0 n) \equiv V_0 \exp(ikn) \\
\vartheta_n &= \vartheta_0 \exp(iKr_0 n) \equiv \vartheta_0 \exp(ikn)
\end{aligned} \tag{20}$$

where the dimensional and dimensionless quasi-wave numbers $K$ and $k$ are introduced. Note that because of the Born-Von Karman condition, the equality $\exp(ikN) = 1$ holds with $N = 6$ for the hexagonal model. Substituting solutions (20) into the system (19), we obtain a system of two equations for amplitudes $v_0$ and $\vartheta_0$:

$$\Omega^2 v_0 = a(k)v_0 + ic(k)\vartheta_0$$
$$\Omega^2 \vartheta_0 = -ic(k)\vartheta_0 + b(k)v_0 \quad (21)$$

where functions $a(k)$, $b(k)$ i $c(k)$ are as follows

$$a(k) = 2[1 + \cos(k) + 9\Lambda(1 + \cos(2k)) + S(1 + \cos(3k))] \quad (22)$$

$$b(k) = 6[1 - \cos(k) + \Lambda(1 - \cos(2k))] \quad (23)$$

$$c(k) = 2\sqrt{3}(\sin(k) + 3\Lambda \sin(2k)) \quad (24)$$

The system of equations (21) is equivalent to the eigenvalue problem for the Hermitian operator $\hat{\mathbf{D}}$:

$$\hat{\mathbf{D}}\begin{pmatrix} v_0 \\ \vartheta_0 \end{pmatrix} = \begin{pmatrix} a(k) & ic(k) \\ -ic(k) & b(k) \end{pmatrix}\begin{pmatrix} v_0 \\ \vartheta_0 \end{pmatrix} = \Omega^2 \begin{pmatrix} v_0 \\ \vartheta_0 \end{pmatrix} \quad (25)$$

The condition for the existence of non-trivial solutions of Eq.(25) is the equality of the determinant of the matrix to zero: $\det\|\hat{\mathbf{D}} - \Omega^2 \hat{\mathbf{I}}\| = 0$, where $\hat{\mathbf{I}}$ is the identity matrix. This condition leads to the characteristic equation:

$$\left(\Omega^2 - a(k)\right)\left(\Omega^2 - b(k)\right) - c^2(k) = 0 \quad (26)$$

This quadratic equation with respect to $\Omega^2$ has two obvious roots:

$$\Omega_\pm^2(k) = \frac{a(k) + b(k)}{2} \pm \sqrt{\left(\frac{a(k) - b(k)}{2}\right)^2 + c^2(k)} \quad (27)$$

As known from crystal lattice theory [10], the "minus" sign in formula (27) corresponds to the acoustic branch of oscillations and the "plus" sign to the optical branch. Their forms depend on the ratios of stiffness coefficients, i.e., on the values of the parameters $\Lambda$ and $S$. In Fig. 4, we show the dependences of the acoustic branch $\Omega_1(k)$ and the optical branch $\Omega_2(k)$ of the dispersion law on the quasi-wave number $k$ for the fixed values of $\Lambda = 0.05$ and $S = 0.116$. The definite choice of the stiffness parameters $\Lambda$ and $S$ is conditioned by their correspondence to the set of model parameters of the graphene hexagon found as a result of the data analysis of crystallographic and spectral experiments in Ref. 11. According to these data, the values and ratios of the force constants of the interactions between the first, second, and third neighbors, measured in units of newton/meter, correspond to the above-chosen values $\Lambda$ and $S$, and they are as the following: $\kappa_1 : \kappa_2 : \kappa_3 = 338 : 50.5 : 19.6$

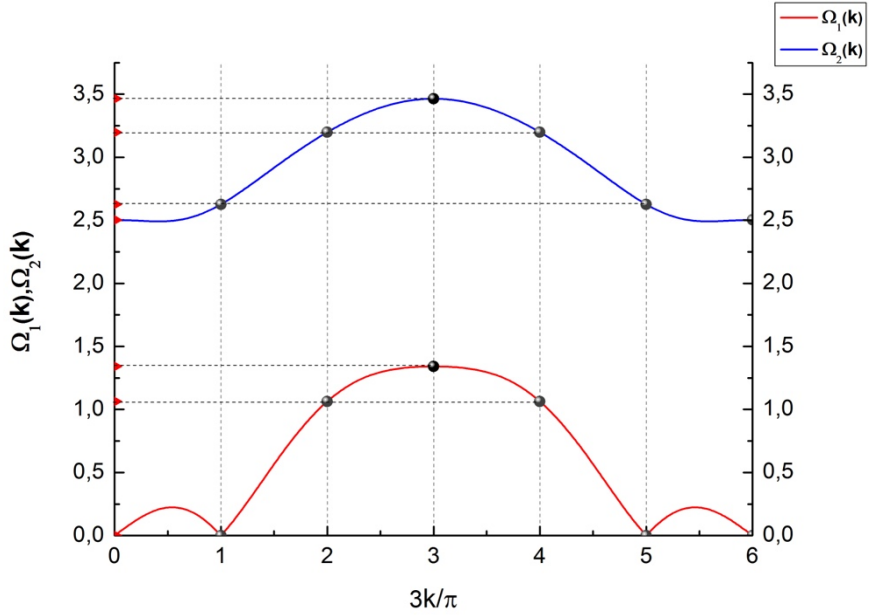

*Fig*. 4. The graphic scheme of finding the discrete frequencies on the acoustic and optical branches of the dispersion law at $\Lambda = 0.05, S = 0.116$.

As seen in Fig. 4, there are two frequency ranges where discrete oscillation frequencies can exist and a frequency gap where the existence of discrete frequencies is prohibited. Exact values of discrete eigenfrequencies are easily found if we use the cyclic Born-Karman condition, which determines the discrete spectrum of allowed values of the quasi-wave number. For a system of six particles, we have

$$k = k_p = \frac{\pi}{3} p, \quad p = 0,1,2,...5. \tag{28}$$

This condition means that the elementary cell in the reciprocal lattice is divided by the allowed values of the quasi-wave number into six equal intervals. Using the branches of the dispersion law, it is easy to identify the discrete frequencies. It is necessary to draw perpendiculars from the points of allowed values of the quasi-wave number, multiples of π/3, to the intersection with the branches of the dispersion law. In Fig. 4, the intersection points of branches of the dispersion law and perpendiculars are denoted by the gray balls. Their ordinates, marked with red triangles on the vertical axis, are equals to the discrete frequencies of the hexagon oscillations.

The eigenvectors $\mathbf{q}_\Omega = (v_1,...v_6, \vartheta_1,...\vartheta_6)$, corresponding to the eigenfrequencies $\Omega$, are given by formulas (20), where instead of the quasi-wave number $k$ its value determined by the Born-Karman condition should be substituted: $v_n = v_0 \exp(ik_p n)$ and $\vartheta_n = \vartheta_0 \exp(ik_p n)$. This means that there are two orthogonal eigenfunctions $\cos(k_p n)$ and $\sin(k_p n)$, a linear combination of which gives the real solutions of the system of Eqs. (19).

Note that this approach for the determination of the discrete spectra allows us to introduce a new classification of frequencies in the molecular spectra based on the found branches of the dispersion law and the known features of oscillations that belong to them. Such an analysis can be effectively applied to more complex cyclic models, while for the model studied in this paper, the expressions for the total spectrum of discrete frequencies are found in an explicit form. Now we focus on qualitative analysis of the frequency spectrum behavior using the evolution of dispersion curves with changing the stiffness parameters.

There are twelve degrees of freedom and twelve eigenfrequencies for six particles in 2D space. We start from zero frequency $\Omega^2 = w_0^2 \equiv 0$. When there is only the first type of springs, i.e., in the approximation of the nearest neighbors, then the acoustic branch degenerates to the zero line, and all six frequencies become zero. The mode softening is said to occur, which usually precedes the system instability. We have already discussed what kind of drastic deformation is possible in this case without changing the energy. The situation is interesting if there are the first and third types of springs, but the second is absent. This choice of parameters corresponds to the setting of experiments with mechanical string models [6,7]. The dispersion curves' behavior is presented in Fig. 5, depending on the parameter $\Lambda$.

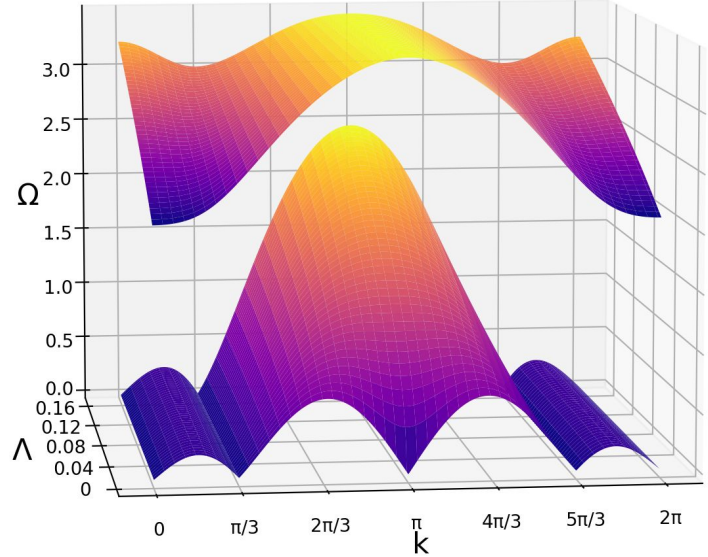

*Fig.* 5. Evolution of the dispersion law branches depending on the parameter $\Lambda$ at $S = 0.116$.

As seen in Fig. 5, at $\Lambda = 0$ the degeneration of the zero frequency is removed at points $k_{2,4} = 2\pi/3, 4\pi/3$, but it remains at $k_3 = \pi$. The further evolution of the curves shows that the zero frequency exists always at three points, $k_0 = 0$ and $k_{1,5} = \pi/3, 5\pi/3$. Below we show the normal modes that responsible for such a frequency behavior. As seen in Fig. 5, the optical frequency range decreases significantly when the parameter $\Lambda$ grows. On the other hand, with growing the parameter $S$ at the fixed $\Lambda$ the qualitative character of the optical branch is the same, but the acoustic curve acquires minimum instead maximum at the edge of the Brillouin zone, as shown in Fig. 6.

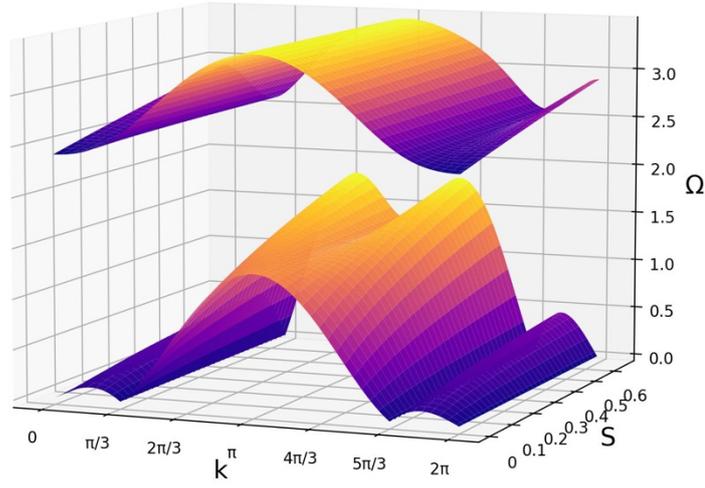

*Fig*. 6. Evolution of the dispersion law branches depending on the parameter $S$ at $\Lambda = 0.05$.

Saying about a frequency degeneration in the experimental spring model, we reveal that two high frequencies can coincide in the experimental case when the springs with the same stiffness coefficient are used to connect the nearest, second and third pair of balls. In Fig. 7, we present the dispersion curves at parameters $\Lambda = 0.3$ and $S = 1.7$, which are close to values $\Lambda_0 = 1/3$ and $S_0 = 2$ corresponding the equality of the stiffness coefficients $\kappa_1 = \kappa_2 = \kappa_3$. The typical situation of the crossover and hybridization of vibrations is observed in this case [10]. The real degeneration of frequencies occurs at the critical values $\Lambda_0$ and $S_0$ in three points $k = 2\pi/3, \pi, 4\pi/3$.

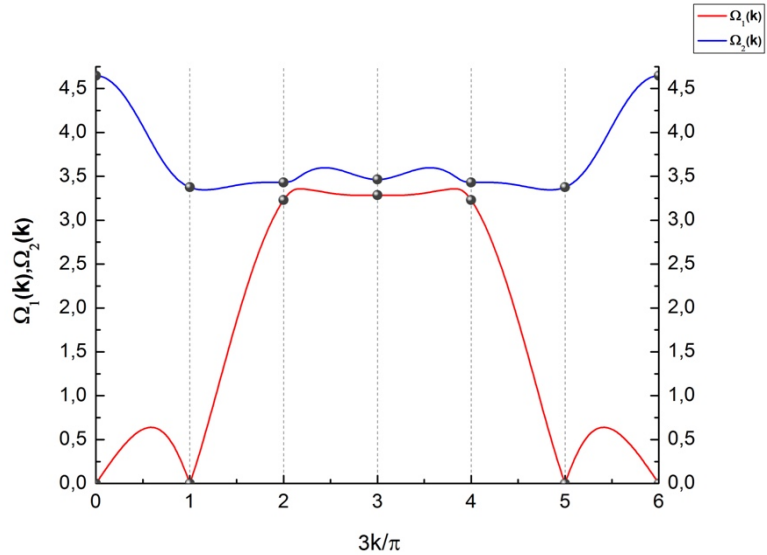

*Fig*. 7. The dispersion curves and the discrete frequencies near the dispersion crossover at $\Lambda = 0.3, S = 1.7$.

The interval of the parameter values in Fig. 5 and Fig. 6 are chosen especially so that the pair $\Lambda = 0.05$ and $S = 0.116$ corresponds to the graphene hexagon parameters, while the pair $\Lambda = 1/6$ and $S = 2/3$ may be suitable for estimating the benzene parameters. Indeed, as follows from the data of Ref. 3 the parameter $\kappa_1$ for "one and a half" bond between carbons can be taken as 735 N/m. We choose a possible value of $\kappa_2$ as half of $\kappa_1$, and the parameter $\kappa_3$ as a third part of $\kappa_1$. One order of magnitude of the decreasing coefficients are argued by the equality of all six main bonds in the benzene ring. In Fig. 8, we show the dispersion curves and discrete frequencies for this set of the benzene hexagon parameters.

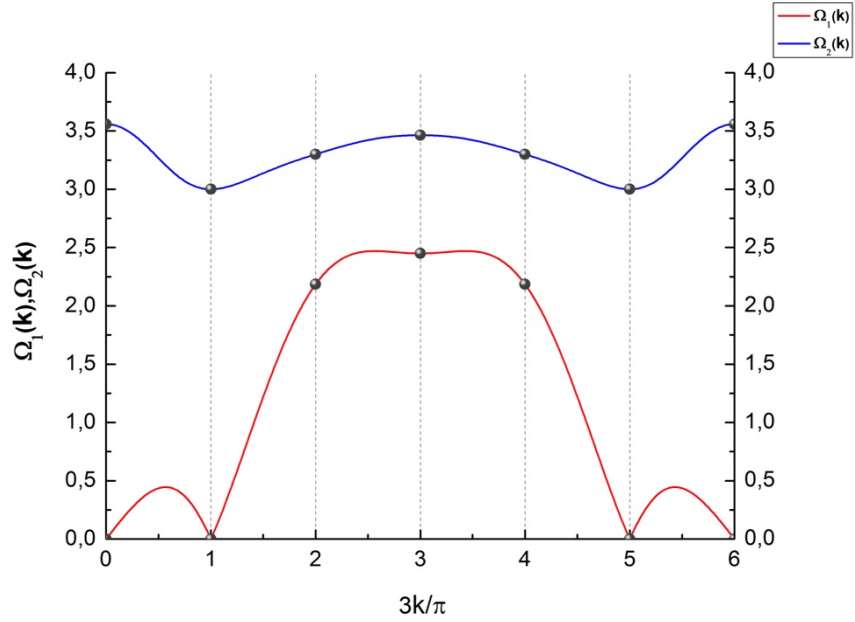

*Fig.* 8. The acoustic and optical branches of the dispersion law at $\Lambda = 1/6$, $S = 2/3$.

Now we present the exact squared frequencies and corresponding solutions of the system (19) as normal modes of the hexagonal model. For this, we substitute the allowed values of the quasi-wave number (28) into dependences (27) of the dispersion law, substituting $k_p$ into formulas (22)–(24) beforehand. As a result, we obtain explicit dependences of the oscillation frequencies on the parameters of the model $\Lambda$ and $S$.

The zero frequency $\Omega^2 = w_0^2 \equiv 0$ corresponds to the normal mode with $p = 0$, $k_0 = 0$, which $\mathbf{q}_1 = (0,0,0,0,0,0;1,1,1,1,1,1)$ and it describes the free angle rotation of the hexagon as a whole around the axis passing through the center of mass. The superposition of two other modes with zero frequency can describe the free angle rotation of the hexagon around the axis passing through any of its particles. These modes are as follows

$$p = 1, \quad k_1 = \frac{\pi}{3}, \qquad \mathbf{q}_2 = \left(0,-1,-1,0,1,1; \frac{1}{\sqrt{3}}(-1,0,2,3,2,0)\right)$$

$$p = 5, \quad k_5 = \frac{5\pi}{3}, \qquad \mathbf{q}_3 = \left(-1,-1,0,1,1,0; \frac{1}{\sqrt{3}}(-2,0,1,0,-2,-3)\right)$$

The squared frequency $\Omega^2 = w_1^2 \equiv 4 + 4\lambda + 2\sigma$ corresponds to $p = 0$, $k_0 = 0$, and its normal mode $\mathbf{q}_4 = (1,1,1,1,1,1;0,0,0,0,0,0)$ describes the breathing of the hexagon, i.e., particles oscillate radially and do not change their angles.

The squared frequency $\Omega^2 = w_2^2 \equiv 6+2\lambda$ corresponds to the normal modes of oppositely moving waves with a length equal to the length of the hexagon ring:

$$p=1, \quad k_1 = \frac{\pi}{3}, \quad \mathbf{q}_5 = \left(1,1,0,-1,-1,0; \frac{1}{\sqrt{3}}(-1,1,2,1,-1,-2)\right)$$

$$p=5, \quad k_5 = \frac{5\pi}{3}, \quad \mathbf{q}_6 = \left(2,1,-1,-2,-1,1; \sqrt{3}(0,1,1,0,-1,-1)\right)$$

The squared frequency $\Omega^2 = w_3^2 \equiv 4\lambda$ depends only on the stiffness of the second-type spring. Its normal mode describes antiphase nearest neighbor oscillations along radii, which do not change the potential energies of springs of the first and third types, but only the second type. It is very important to reveal the frequency of this mode in the spectral experiments with benzene in order to estimate the strength of second neighbor interactions in the molecule. The mode is the following:

$$p=3, \quad k_3 = \pi, \quad \mathbf{q}_7 = (-1,1,-1,1,-1,1;0,0,0,0,0,0)$$

The squared frequency $\Omega^2 = w_4^2 \equiv 12$ depends only on the stiffness of the first-type spring. Its normal mode describes antiphase nearest neighbor oscillations along angle and does not change the lengths of springs of the second and third types. The revealing of this mode in the spectral experiment directly gives the estimate for the strength of the main carbon-carbon interaction. The mode looks like

$$p=3, \quad k_3 = \pi, \quad \mathbf{q}_8 = (0,0,0,0,0,0;1,-1,1,-1,1,-1;)$$

It is remarkable that the ratio of amplitudes $v_0/\vartheta_0$ for all the above modes does not depend on the stiffness parameters of the hexagon. For the last four eigenfrequencies, we need to calculate special coefficients depending on them. We introduce the following coefficients

$$C_1 = \sqrt{3} \cdot \frac{4-\sigma + \sqrt{(3-\lambda)^2 + (4-\sigma)^2}}{3-\lambda}, \quad C_2 = -\sqrt{3} \cdot \frac{4-\sigma - \sqrt{(3-\lambda)^2 + (4-\sigma)^2}}{3-\lambda}$$

We note that the inequalities $\lambda < 3$ and $h < 4$, which mean that $\kappa_1 > \kappa_2$ and $\kappa_1 > \kappa_3$ ensure that coefficients $C_1 > 0, C_2 > 0$.

The squared frequency $\Omega^2 = w_6^2 \equiv 5+\lambda+\sigma - \sqrt{(3-\lambda)^2 + (4-\sigma)^2}$ corresponds to the normal modes with the wavelength as half of the hexagon ring length and relates to the acoustic dispersion curve

$$p=2, \quad k_2 = \frac{2\pi}{3}, \quad \mathbf{q}_9 = (C_1(-1,1,0,-1,1,0);1,1,-2,1,1,-2)$$

$$p=4, \quad k_4 = \frac{4\pi}{3}, \quad \mathbf{q}_{10} = (-2,1,1,-2,1,1; C_2(0,1,-1,0,1,-1))$$

The squared frequency $\Omega^2 = w_6^2 \equiv 5+\lambda+\sigma + \sqrt{(3-\lambda)^2 + (4-\sigma)^2}$ corresponds to the normal modes with the wavelength as half of the hexagon ring length and relates to the optical dispersion curve

$$p = 2, \quad k_2 = \frac{2\pi}{3}, \quad \mathbf{q}_{11} = (C_2(1,-1,0,1,-1,0);1,1,-2,1,1,-2)$$

$$p = 4, \quad k_4 = \frac{4\pi}{3}, \quad \mathbf{q}_{12} = (-2,1,1,-2,1,1; C_1(0,-1,1,0,-1,1))$$

Concluding this section, we note that the values of the parameters $\Lambda$ and $S$, which are characteristic of graphene hexagons, chosen for plotting the graphs, are relatively small, while for the benzene molecule, deuterated benzene, and substituted benzenes with halogens, these parameters are much larger. The behavior of vibration frequencies in this range of parameters will be discussed in the next section. But for comparison with benzene-like molecules, one should, first of all, take into account the presence of hydrogen, deuterium, or halogens in the molecule and, accordingly, their influence on the dynamics of the hexagonal model, which is done in the next section.

**Vibration spectra of benzene-like models accounting motion of hydrogen, deuterium, and halogens**

The mechanical string model of a benzene-like molecule includes, in addition to six particles with mass $m_1$ as carbon analogues, six new balls with mass $m_2$ as hydrogen analogues, each of which forms a pair with one carbon, as shown in Fig. 2. The coupling between the balls is provided by springs of three types as discussed in previous sections. We denote the stiffness coefficient of the new springs as $\kappa_4$.

The positions of new particles are described by cylindrical coordinates $b_n$, $\chi_n$ and $z_n^H$. At rest, hydrogen is at a distance $a_0$ from carbon and at a distance $b_n^{(0)} = r_0 + a_0$ from the origin. It lies on the same line as carbon, so the equilibrium angle is $\chi_n^{(0)} = \varphi_n^{(0)}$, and its coordinate is $z_n^H = 0$. We number hydrogen atoms by analogy with carbon atoms. The chart showing the mutual deviation of carbon and hydrogen atoms is shown in Fig. 9.

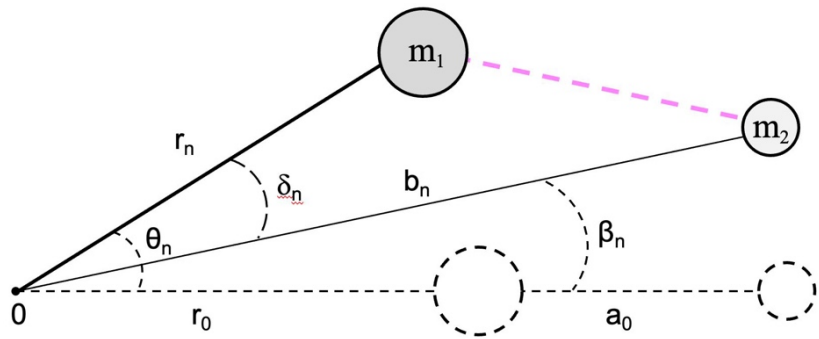

*Fig.* 9. Scheme for calculating the length change in the spring with stiffness coefficient $\kappa_4$ during particle oscillations.

During small oscillations, particles $m_1$ are displaced according to formulas (2), i.e., $r_n = r_0 + u_n$ and $u_n \ll r_0$, and their angle deviations $\theta_n$ are small. The contribution of the coordinate $z_n$ to the spring length change is a nonlinear effect. We denote the displacement of particle $m_2$ along the radius by $h_n$, and the deviation angle from the equilibrium position as $\beta_n = \chi_n - \chi_n^{(0)}$. Then the radius $b_n$ and angle difference $\delta_n$ in Fig. 9 are expressed as the following

$$b_n = r_0 + a_0 + h_n, \qquad \delta_n = \beta_n - \theta_n \qquad (30)$$

Using the scheme presented in Fig. 9, we calculate the change in the length of the new spring (the magenta dashed line) during the oscillations of the benzene-like model. According to the law of cosines, the new distance between the particles $m_1$ and $m_2$ is equal

$$l_n = \sqrt{r_n^2 + b_n^2 - 2b_n r_n \cos(\delta_n)} \qquad (31)$$

At small angles $\theta_n$ and $\beta_n$ their difference $\delta_n$ is also small and $\cos(\delta_n) \approx 1 - \delta_n^2/2$. Thus, in the harmonic approximation, the cosine value is taken equal to unity, hence, $l_n = b_n - r_n$. Substituting $r_n = r_0 + u_n$ and $b_n = r_0 + a_0 + h_n$ into formula (31), we obtain a simple expression for the spring length and its change

$$\psi_n = l_n - a_0 = h_n - u_n \equiv r_0(g_n - v_n) \qquad (32)$$

where the dimensionless displacement $g_n$ of the particle $m_2$ is introduced. The potential energy of Hooke's law interaction between two particles is a quadratic function of $\psi_n$:

$$U_4 = \frac{\kappa_4}{2}\sum_{n=1}^{6}\psi_n^2 = \frac{\kappa_4 r_0^2}{2}\sum_{n=1}^{6}(g_n - v_n)^2 \qquad (33)$$

Note that the present spring model actually takes into account the angle interaction, usually interpreted as non-central, through the cosine function in formula (31), which ensures the equilibrium position of the particle $m_2$ exactly on the same line as the particle $m_1$. However, as seen in Fig. 9, the particle oscillations measured by the angle $\delta_n$, as well as by the coordinates $z_n^H$ are *the anharmonic effects* in this model, as it takes place in the case of $z_n$-coordinate in the hexagon model. The kinetic energy of particles with mass $m_2$ is the following:

$$T_H = \frac{m_2}{2}\sum_{n=1}^{6}\left(\frac{dh_n}{dt}\right)^2 + \frac{I_\beta}{2}\sum_{n=1}^{6}\left(\frac{d\beta_n}{dt}\right)^2 + \frac{m_2}{2}\sum_{n=1}^{6}\left(\frac{dz_n^H}{dt}\right)^2 \qquad (34)$$

where $I_\beta = m_2(r_0 + a)^2$ is the moment of inertia of atoms. However, the last two terms, which determine the dynamics of coordinates $\beta_n$ and $z_n^H$ should be considered outside the harmonic approximation, namely, within the framework of nonlinear equations. Thus, only the first term in the form

$$T_h = \frac{m_2 r_0^2}{2}\sum_{n=1}^{6}\left(\frac{dg_n}{dt}\right)^2 \qquad (35)$$

should be added to the kinetic energy of the carbon hexagon and the term $U_4$, given by Eq. (33), to the hexagon potential energy $U$. Finally, for the mechanical spring model of a benzene-like molecule, we obtain the total Lagrangian of the following form:

$$\tilde{L} = \frac{E_0}{2}\left( \sum_{n=1}^{6}\left(\frac{dv_n}{d\tau}\right)^2 + \sum_{n=1}^{6}\left(\frac{d\theta_n}{d\tau}\right)^2 + \gamma\sum_{n=1}^{6}\left(\frac{dg_n}{d\tau}\right)^2 - \sum_{n=1}^{6}(v_{n+1}+v_n+\sqrt{3}\theta_{n+1}-\sqrt{3}\theta_n)^2 - \right.$$

$$\left. -\Lambda\sum_{n=1}^{6}(3v_{n+2}+3v_n+\sqrt{3}\theta_{n+2}-\sqrt{3}\theta_n)^2 - S\sum_{n=1}^{6}(v_{n+3}+v_n)^2 - Q\sum_{n=1}^{6}(g_n-v_n)^2 \right)$$

(36)

where the following parameters for the ratio of masses and the new stiffness coefficient ratio are introduced as follows

$$\gamma \equiv \frac{1}{\rho} = \frac{m_2}{m_1}, \qquad Q = \frac{4\kappa_4}{\kappa_1} \tag{37}$$

Now we derive the Lagrange equations for the generalized model, described by the Lagrangian (36), and find their solutions. After appropriate differentiation of the Lagrangian (36) with respect to the generalized coordinates and their velocities, we obtain three equations:

$$\frac{d^2 v_n}{d\tau^2} - \frac{1}{E_0}\frac{\partial \tilde{L}}{\partial v_n} = 0, \qquad \frac{d^2 \theta_n}{d\tau^2} - \frac{1}{E_0}\frac{\partial \tilde{L}}{\partial \theta_n} = 0, \qquad \frac{d^2 g_n}{d\tau^2} - \frac{1}{E_0}\frac{\partial \tilde{L}}{\partial g_n} = 0. \tag{38}$$

We seek for solutions of system (38), as before, in the form of stationary oscillations

$$v_n = \nu_n e^{-i\Omega\tau}, \qquad \theta_n = \vartheta_n e^{-i\Omega\tau}, \qquad g_n = \mu_n e^{-i\Omega\tau}. \tag{39}$$

Substituting (39) into (38), we obtain a system of homogeneous algebraic equations with constant coefficients:

$$\Omega^2 \nu_n = 2\nu_n + \nu_{n+1} + \nu_{n-1} + \sqrt{3}(\vartheta_{n+1}-\vartheta_{n-1}) +$$
$$+ 3\Lambda(6\nu_n + 3\nu_{n+2} + 3\nu_{n-2} + \sqrt{3}(\vartheta_{n+2}-\vartheta_{n-2})) +$$
$$+ S(2\nu_n + \nu_{n+3} + \nu_{n-3}) + Q(\nu_n - g_n)$$
$$\Omega^2 \vartheta_n = 6\vartheta_n - 3\vartheta_{n+1} - 3\vartheta_{n-1} + \sqrt{3}(\nu_{n-1}-\nu_{n+1}) +$$
$$+ 3\Lambda(2\vartheta_n - \vartheta_{n+2} - \vartheta_{n-2} + \sqrt{3}(\nu_{n-2}-\nu_{n+2}))$$
$$\gamma\Omega^2 \mu_n = Q(\mu_n - \nu_n) \tag{40}$$

Due to the cyclic conditions, the solutions of the system of Eqs. (40) are found in the explicit form [10]:

$$\nu_n = \nu_0 \exp(ikn), \qquad \vartheta_n = \vartheta_0 \exp(ikn), \qquad \mu_n = \mu_0 \exp(ikn). \tag{41}$$

Substituting solutions (41) into the system (40), we obtain a system of three equations for the amplitudes $\nu_0$, $\vartheta_0$ and $\mu_0$, which we write in such a sequence:

$$\gamma\Omega^2 g_0 = Qg_0 - Qv_0$$
$$\Omega^2\theta_0 = b(k)\theta_0 - ic(k)v_0 \quad (42)$$
$$\Omega^2 v_0 = ic(k)\theta_0 + \tilde{a}(k)v_0 - Qg_0$$

When writing the equations (42), we have used the functions $b(k)$ and $c(k)$ from Eqs. (23) and (24) and introduced the function $\tilde{a}(k) = a(k) + Q$, where the function $a(k)$ is defined by formula (22). The condition for the existence of non-trivial solutions of the system of equations (42) is the equality of the determinant of the following matrix to zero:

$$\det \begin{Vmatrix} Q - \gamma\Omega^2 & 0 & -Q \\ 0 & b(k) - \Omega^2 & -ic(k) \\ -Q & ic(k) & \tilde{a}(k) - \Omega^2 \end{Vmatrix} = 0 \quad (43)$$

Expanding the determinant, we get the following cubic equation for the squared frequency $\Omega^2$:

$$\left(Q - \gamma\Omega^2\right)\left\{\left(b(k) - \Omega^2\right)\left(\tilde{a}(k) - \Omega^2\right) - c^2(k)\right\} - Q^2\left(b(k) - \Omega^2\right) = 0 \quad (44)$$

It is easy to see that this equation is equivalent to the condition that the determinant of some Hermitian matrix is equal to zero

$$\det \begin{Vmatrix} \dfrac{Q}{\gamma} - \Omega^2 & 0 & -\dfrac{Q}{\sqrt{\gamma}} \\ 0 & b(k) - \Omega^2 & -ic(k) \\ -\dfrac{Q}{\sqrt{\gamma}} & ic(k) & \tilde{a}(k) - \Omega^2 \end{Vmatrix} \equiv \det\left\|\hat{\mathbf{L}} - \Omega^2\hat{\mathbf{I}}\right\| = 0 \quad (45)$$

Thus, the characteristic equation (45) is equivalent to the eigenvalue problem for the Hermitian operator $\hat{\mathbf{L}}$ and the squared frequencies are nothing but its eigenvalues, which are necessarily real quantities. This means that the cubic equation (44) has all three real roots. Thus, there exist necessarily three branches of the dispersion law, i.e., three dependencies $\Omega^2_{1,2,3}(k)$.

The characteristic cubic equation (44) can be rewritten in the standard form:

$$\Omega^6 + \tilde{A}(k)\Omega^4 + \tilde{B}(k)\Omega^2 + \tilde{C}(k) = 0 \quad (46)$$

where the coefficients are the following functions:

$$\tilde{A}(k) = -(\tilde{a}(k) + b(k) + \rho Q) \quad (47)$$
$$\tilde{B}(k) = \tilde{a}(k)b(k) - c^2(k) + \rho Q(\tilde{a}(k) + b(k) - Q)$$
$$\tilde{C}(k) = \rho Q\left(b(k)(Q - \tilde{a}(k)) + c(k)^2\right)$$

Here and further, it is convenient to use the denotation $\rho$ instead $1/\gamma$.

Now we present three real roots of this equation (46) as cumbersome but explicit functions of the quasi-wave number using Cardano's formulas [12]. First, as usual, we need to reduce the equation (46) to the depressed cubic

$$y^3 + p(k)y + g(k) = 0 \tag{48}$$

by replacing $y = \Omega^2 + \dfrac{\tilde{A}(k)}{3}$ and introducing the coefficients as follows:

$$p(k) = -\left(\dfrac{\tilde{A}(k)^2}{3} - \tilde{B}(k)\right) < 0, \tag{49}$$

$$g(k) = 2\left(\dfrac{\tilde{A}(k)}{3}\right)^3 - \dfrac{\tilde{A}(k)\tilde{B}(k)}{3} + \tilde{C}(k). \tag{50}$$

Whereupon, the squared frequencies $\Omega^2$, as the roots of the cubic equation (46), are written explicitly presenting three branches of the dispersion law

$$\Omega_1^2(k) = -\dfrac{\tilde{A}(k)}{3} + 2\sqrt{\dfrac{|p(k)|}{3}}\cos\left(\dfrac{\alpha(k)}{3}\right) \tag{51}$$

$$\Omega_{2,3}^2(k) = -\dfrac{\tilde{A}(k)}{3} - 2\sqrt{\dfrac{|p(k)|}{3}}\cos\left(\dfrac{\alpha(k)}{3} \pm \dfrac{\pi}{3}\right) \tag{52}$$

where the dependence of the function $\alpha(k)$ is as follows:

$$\alpha(k) = \arccos\left(-\dfrac{g(k)}{2}\sqrt{\left(\dfrac{3}{|p(k)|}\right)^3}\right). \tag{53}$$

Below we demonstrate graphically the obtained branches for specific parameter values $\Lambda$, $S$, $Q$. The studied benzene-like model is believed to be applicable to the description of the benzene molecule as well as fully deuterated and halogenated benzenes. In Ref. 3, the values of the parameters of interactions between carbon and hydrogen, deuterium, fluorine, chlorine, bromine, and iodine are presented. In Fig. 10, we use corresponding the mass ratio $\gamma$ and the stiffness ratio $Q$, and parameters $\Lambda$ and $S$ of the hexagon model (Fig. 8) to plot graphs for pairs (a) C–H, (b) C–D, (c) C–F, (d) C–Cl, (e) C–Br, and (f) C–I. From (a) to (f), the parameters $\gamma$ and $\kappa_4$ in units N/m change respectively as follows: $\gamma = 1/12$, 1/6, 1.58, 2.96, 6.58, 10.58 and $\kappa_4 = $ 490, 490, 596, 364, 313, 265. We see that in the case of benzene (Fig. 10a), the acoustic and optical branches are practically the same as in the hexagon model in Fig. 8 because the hydrogen branch is much above and does not affect them. Thus the dispersion curves indicate the well-known separation of the benzene frequency spectrum into three groups. The isotope effect with deuterium is presented in Fig. 10b. The interaction between the branches begins already from fluorine and strengthens with increasing $\gamma$. The typical crossover situation occurs between a previously acoustic branch and the practically local frequency of the carbon-halogen oscillation, which is seen as the red flat band portion in the case of bromine and especially iodine (Fig. 10e,f). Thus, the degeneration of frequencies is always removed, and the oscillation hybridization takes place.

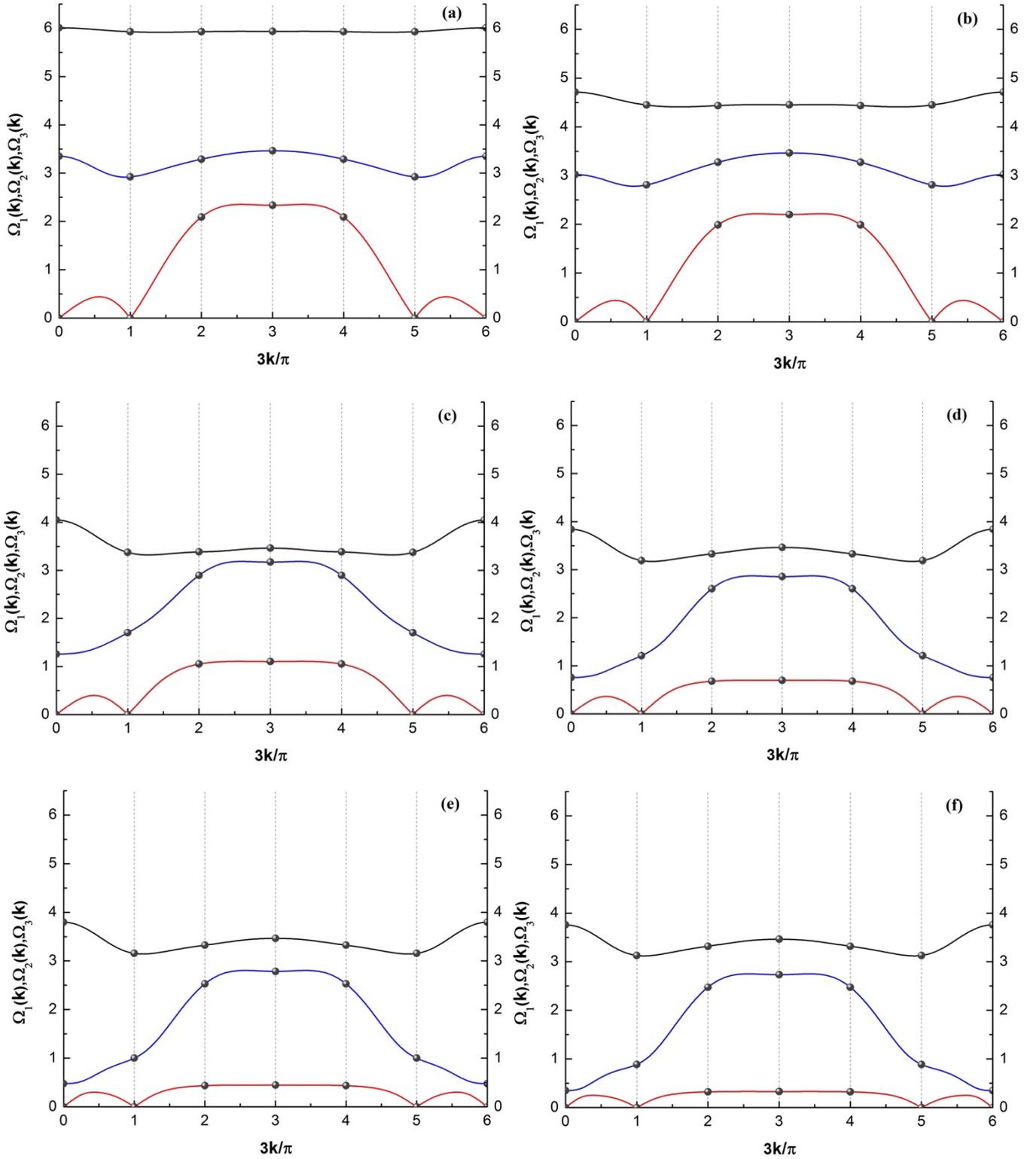

*Fig.* 10. The dispersion law branches and the discrete frequency spectra of benzene with (a) hydrogen and (b) deuterium, and fully substituted benzenes with (c) fluorine, (d) chlorine, (e) bromine, (f) iodine, respectively.

It is remarkable that the discrete frequency spectrum, marked by the grey balls in Fig. 10, can be found exactly as the explicit functions of the benzene-like model parameters. The branch dependences of the dispersion law allows us to determine a discrete set of vibration frequencies of a benzene-like molecule by substitution in them the values of the quasi-wave number, satisfying the Born-Karman conditions (28). When $p = 0$ and the quasi-wave number $k_0 = 0$, then the functions in the cubic equation (44) are significantly simplified and take the form:

$$\tilde{a}(0) = w_1^2 + Q, \qquad b(0) = c(0) = 0, \tag{54}$$

In this case, the characteristic equation (46) has one zero root $\Omega^2 = \Omega_1^2(0) = 0$ and actually degenerates to the quadratic:

$$\Omega^2\left(\Omega^4 - \overline{B}(0)\Omega^2 + \overline{C}(0)\right) = 0 \tag{55}$$

where the coefficients are equal as

$$\overline{B}(0) \equiv -\tilde{A}(0) = w_1^2 + (1+\rho)Q, \qquad \overline{C}(0) \equiv \tilde{B}(0) = \rho Q w_1^2, \tag{56}$$

Then, we solve the quadratic equation for $\Omega^2$ in parentheses of Eq.(55), and after elementary transformations of its roots, we find squared frequencies in the explicit form:

$$\Omega_{2,3}^2(0) = \frac{1}{2}\left\{w_1^2 + (1+\rho)Q \pm \sqrt{\left(w_1^2 - (1+\rho)Q\right)^2 + 4Q w_1^2}\right\}. \tag{57}$$

Both frequencies are placed on the ordinate axis in Fig. 10.

When $p = 1$ and the quasi-wave number $k = \pi/3$, then the functions in the cubic equation (44) are expressed directly through half of the squared frequency $w_2^2 = 2(3 + \lambda)$:

$$\tilde{a}\left(\frac{\pi}{3}\right) = \frac{1}{2}w_2^2 + Q, \qquad b\left(\frac{\pi}{3}\right) = c\left(\frac{\pi}{3}\right) = \frac{1}{2}w_2^2. \tag{58}$$

The characteristic equation (46) again has one zero root $\Omega^2 = \Omega_1^2\left(\frac{\pi}{3}\right) = 0$ and after its exclusion, degenerates to the quadratic:

$$\Omega^4 - \overline{B}\left(\frac{\pi}{3}\right)\Omega^2 + \overline{C}\left(\frac{\pi}{3}\right) = 0 \tag{59}$$

where the coefficients are equal to

$$\overline{B}\left(\frac{\pi}{3}\right) = -\tilde{A}\left(\frac{\pi}{3}\right) = w_2^2 + (1+\rho)Q, \qquad \overline{C}\left(\frac{\pi}{3}\right) = \left(\rho + \frac{1}{2}\right)Q w_2^2. \tag{60}$$

The two roots of a quadratic equation are as follows:

$$\Omega_{2,3}^2\left(\frac{\pi}{3}\right) = \frac{1}{2}\left\{w_2^2 + (1+\rho)Q \pm \sqrt{\left(w_2^2 - (1+\rho)Q\right)^2 + 2Q w_2^2}\right\} \tag{61}$$

When $p = 3$ and $k = \pi$ the functions in the cubic equation (44) become as follows:

$$\tilde{a}(\pi) = w_3^2 + Q, \qquad b(\pi) = 12, \qquad c(\pi) = 0. \tag{62}$$

Since the last coefficient in (62) is zero, equation (44) is rewritten as

$$\left(\Omega^2 - b(\pi)\right)\left\{\left(\gamma\Omega^2 - Q\right)\left(\Omega^2 - \tilde{a}(\pi)\right) - Q^2\right\} = 0 \qquad (63)$$

From Eq.(63), it follows that the root $\Omega_1^2(\pi) = 12$ is independent of stiffness coefficient ratios and remains unchanged even when we add hydrogen, deuterium, or halogens to the hexagon model. It can be simply explained because the normal mode does not change any radial distance between particles. The quadratic equation for the remaining two roots can be rewritten as

$$\left(\Omega^2 - \rho Q\right)\left(\Omega^2 - \tilde{a}(\pi)\right) - \rho Q^2 = 0 \qquad (64)$$

The quadratic Eq.(64) has two standard roots

$$\Omega_{2,3}^2(\pi) = \frac{1}{2}\left\{w_3^2 + (1+\rho)Q \pm \sqrt{\left(w_3^2 + (1-\rho)Q\right)^2 + 4\rho Q^2}\right\} \qquad (65)$$

And only in the cases, when $k = k_2 = 2\pi/3$ and $k = k_4 = 4\pi/3$, Eqs. (44) and (46) still remain truly cubic, and their roots have to be determined by general Cardano's formulas with the substitution of the values $k_2$ and $k_4$ into all three branches of the dispersion law (51)–(53).

Thus, we have found explicitly the vibration spectrum of the mechanical benzene-like model with Hooke's law spring interaction. Its dynamics in the harmonic approximation maintains the specific features of oscillation motion of carbon-based cyclic hexagonal systems, in particular, deuterated and halogenated benzene molecules. As a theoretical perspective, it seems interesting to study the dynamics of stable flat cyclic models with an arbitrary finite number of particles coupled by various kinds of springs in the framework of the approach proposed in the present work. The principal formula (9) for $j$-th spring length change is easy to generalize to the case of arbitrary particle number $N$:

$$l_{n,n+j} - l_{n,n+j}^{(0)} = \sqrt{r_{n+j}^2 + r_n^2 - 2r_{n+1}r_n \cos\left(\frac{2\pi j}{N} + \theta_{n+j} - \theta_n\right)} - l_{n,n+j}^{(0)}, \qquad j = 1,2,3... \qquad (66)$$

Evidently, after linearization near the equilibrium values of the polar coordinates of particles, the length spring changes can be used to derive the Lagrangian and the Lagrange equations and eventually to find the discrete frequency spectrum of harmonic oscillations.

## Conclusion

The main findings of this study are as follows:

(1) The oscillatory dynamics of the hexagonal spring-ball model of benzene-like nanosystems has been analytically investigated in the harmonic approximation. The model captures the main vibrational properties of cyclic nanostructures which include the graphene and carbon nanotube hexagons, and the benzene molecule itself and fully deuterated and halogenated benzenes. This mechanical hexagon is built from balls connected by springs, which ensure interactions with the nearest, second and third neighbors, according to Hooke's law, and stabilize the system in a flat configuration. The geometrical approach has been used to find the spring length change during oscillations, which has been the key point for constructing the Lagrangian of the hexagonal system.

(2) The second clue moment has been a choice of cylindrical coordinates for the description of oscillations of a benzene-like ring, whose Lagrange equations with the cyclic Born-Von Karman condition turned out to be similar to those of one-dimensional two-component (scalar) lattice model. The acoustic and optical branches of the dispersion law of such a lattice model have been found analytically. The imposition of the cyclic Born-Von Karman boundary condition on the stationary solutions of the Lagrange equations establishes the discrete frequency positions on the found branches of the dispersion law. Thus, discrete frequencies can be classified according to their belonging to the branches of the dispersion law. At the same time the shape of the branches determines the existence area of groups of discrete frequencies. As a result, the vibration spectrum of hexagon harmonic oscillations and its normal modes have been exactly found in the form of their explicit dependencies on all model stiffness parameters.

(3) The generalized benzene-like model has been studied, which takes into account the presence of hydrogen, deuterium, and halogens in the benzene molecule itself and the substituted benzenes, respectively. The analytical description of the associated lattice model has been carried out, resulting in finding the acoustic and two optical branches of the dispersion law. The effect of the hybridization of vibration modes and pushing of spectral branches in a crossover situation is revealed. The branches are hybridized with forming the forbidden frequency gap, while one of the branches behaves like the flat band of the attached impurity [13]. For the generalized model, the discrete frequency dependences on all interaction parameters and the mass ratio of carbon and attached hydrogen, deuterium, and halogens have been found. We also shortly discussed how the obtained results could be used for the analysis of experiments in molecular spectroscopy and generalized for the theoretical description of cyclic systems with an arbitrary finite number of particles.

ACKNOWLEDGMENTS

This research was funded in part by the Luxembourg National Research Fund (FNR) Grant No. 17568826 (ELEMENT). For the purpose of open access, O.C. has applied a Creative Commons Attribution 4.0 International (CC BY 4.0) license to any Author Accepted Manuscript version arising from this submission.